\pdfoutput=1

\documentclass[twocolumn,letterpaper]{article}
\usepackage{times}

\usepackage{graphicx}
\usepackage{url}
\usepackage{mdwlist}
\usepackage{ragged2e}
\usepackage[super]{nth}
\usepackage{color}
\usepackage{soul}
\usepackage{multirow}
\usepackage{fancyhdr}

\usepackage[caption=false,font=footnotesize]{subfig}
\usepackage{floatrow}
\usepackage{setspace}
\newcommand{\drop}[1]{\textcolor{blue}{\st{#1}}}
\renewcommand{\drop}[1]{}

\def\smallerspacecaption{\vspace{-2pt}}

\usepackage[letter]{aspdac2e}

\begin{document}
\date{}

\title{\Large\textbf{Concerted Wire Lifting: Enabling Secure and
	Cost-Effective
	Split Manufacturing}}

\author{\normalsize
 \begin{tabular}[t]{c@{\extracolsep{3em}}c@{\extracolsep{2em}}c@{\extracolsep{2em}}c}
  \large Satwik Patnaik		& \large Johann Knechtel & \large Mohammed Ashraf & \large Ozgur Sinanoglu \\
  \\
  Tandon School of Engineering	& \multicolumn{3}{c}{Design for Excellence Lab, Division of Engineering} \\
  New York University, USA	& \multicolumn{3}{c}{New York University Abu Dhabi, UAE} \\
  sp4012@nyu.edu & \multicolumn{3}{c}{\{johann, ma199, ozgursin\}@nyu.edu} \\
\end{tabular}}
\maketitle

\thispagestyle{myheadings}
\markright{
\copyright~IEEE, 2018, ASPDAC.
	This is the author's version of the work. It is posted here for your personal use;
       not for redistribution.
}
\pagestyle{empty}

{\small\textbf{Abstract---Here we advance the protection of split manufacturing (SM)-based layouts through the
judicious and well-controlled handling of interconnects.
Initially, we explore the cost-security trade-offs of SM,
	which are limiting its adoption.
Aiming to resolve this issue,
       we propose effective and efficient strategies to lift nets to the BEOL.
Towards this end, we design custom ``elevating cells'' which we also provide to the community.
Further, we define and promote a new metric, Percentage of Netlist Recovery (PNR), which can quantify the resilience against gate-level theft of intellectual property (IP) in a manner more meaningful than established metrics.
Our extensive experiments show that we outperform the recent protection schemes regarding security.
For example, we reduce the correct connection rate to
		0\% for commonly considered benchmarks, which is a first in the literature.
Besides, we induce reasonably low and controllable overheads on power, performance, and area (PPA).
At the same time, we also help to lower the commercial cost incurred by SM.
}}

\renewcommand{\arraystretch}{1.07}

\section{Introduction}
\label{sec:introduction}

Nowadays, chip manufacturing is a complex and costly process, where more often than not third-party facilities are involved.
As a result, protecting intellectual property (IP) as well as ensuring trust in the chips becomes challenging.

\begin{figure}[tb]
\centering
\includegraphics[width=.95\columnwidth]{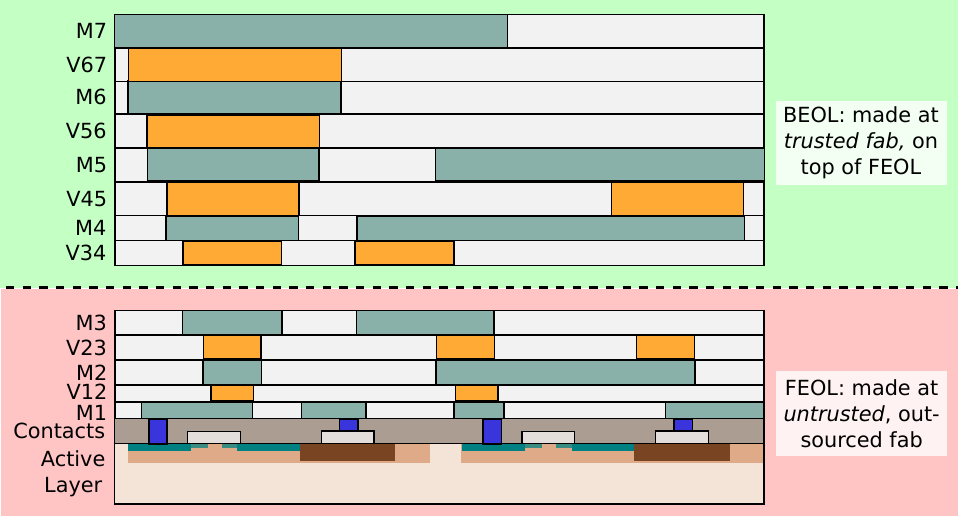}
\caption{Concept of split manufacturing, i.e., the separation of a layout into the FEOL and BEOL parts.
	Note the different pitches across the metal layers.
	As the FEOL part is outsourced, it may require additional protection (such as
	placement perturbation or
	lifting of wires) against fab-based adversaries.
\label{fig:SM_concept}
}
\end{figure}

The IARPA agency
proposed \emph{split manufacturing (SM)}
as a protection technique 
to ward off threats like IP piracy, unauthorized overproduction,
and insertion of hardware Trojans~\cite{mccants11}.
In the simplest embodiment of SM,
the FEOL
is handled by a high-end, competitive off-shore fab which is potentially \emph{untrusted}, while the BEOL is manufactured subsequently at a low-end,
\emph{trusted} facility (Fig.~\ref{fig:SM_concept}).
Hill \emph{et al.\ }\cite{hill13}
successfully
demonstrated the viability of SM by fabricating a 1.3 million-transistor asynchronous FPGA.
Further studies also bear testament to the applicability of SM~\cite{vaidyanathan14_2,bi2015beyond,vaidyanathan14}.
However, the overall acceptance of SM remains behind expectations so far, mainly due to concerns about cost.

The protection offered by SM is based on the fact that 
the FEOL fab does not have access to the complete design, and an attacker \emph{may} thus be hindered
from malicious activities.
The threat models for SM~\cite{garg2017split} are accordingly focused on FEOL-based adversaries which either seek
	to ($i$) retrieve the
	design and/or its IP,
	or ($ii$) insert hardware Trojans.
Some studies also consider both at the same time~\cite{xiao15,glsvlsi2017}.
Here, we address ($i$).

Prior art suggests splitting
after M1,
as such a scenario forces an attacker to tackle a ``vast sea of gates''
with only a few transistor-level interconnects provided along~\cite{vaidyanathan14_2}.
Although splitting after M1 arguably provides the best protection, it also necessitates a high-end BEOL fab for trusted fabrication of all remaining metal layers, including
	the lower layers with very small pitches.  Since this requirement 
may be
considered too costly,
	   some studies propose to split after M4~\cite{xiao15,magana16,magana17}.\footnote{We advocate the terminology ``to split after'' instead of the commonly used ``to split at.'' For example, ``to split at M2'' remains ambiguous whether
	M2 is still within the FEOL or already in the BEOL.  Further, the same uncertainty applies to the vias of V12 and V23, i.e., those between M1/M2 and M2/M3, respectively.  Our
		definition for ``to split after M2'' is that M2 and V12 are still in the FEOL, while the vias of V23 are already in the BEOL.}
	   However, doing so can
undermine security by revealing more structural connectivity information
to an attacker~\cite{rajendran13_split,wang16_sm,sengupta17}.

The key challenge for SM is thus: \emph{how to render split manufacturing practical regarding both security and cost?}

   Here, we address this challenge with a secure and effective approach for SM.
Our key principle is to lift wires to the BEOL in a controlled and concerted manner, considering both cost and security.
Our work can be summarized as follows:
\begin{itemize*}

\item Initially, we revisit the cost-security trade-offs for SM. We explore the prospects of wire lifting and find that naive lifting to higher metal layers can improve
	the security
		albeit at high layout cost.
		Thus, we proclaim the need for cost- and security-aware, concerted lifting schemes.

\item We put forward multiple strategies to select and lift nets.
The key ideas
to achieve strong protection
are ($i$)~to increase the number of protected/lifted nets and ($ii$)~to dissolve hints of physical proximity for those nets.

\item
Based on our strategies, we propose a method for the \emph{concerted lifting of wires}
with controllable impact on power, performance, and area (PPA).
Since we lift wires to higher metal layers (M6, without loss of generality),
our method also helps to lower the commercial cost of SM.

\item
For the actual layout-level lifting,
we design custom ``elevating cells.''
Unlike the prior art,
	our techniques allow to lift and route wires in a controlled manner in the BEOL.

\item We promote a new metric, \emph{Percentage of Netlist Recovery (PNR)},
which quantifies the resilience offered by any SM protection scheme
against varyingly effective attacks.

\item We conduct a thorough evaluation of layout cost and security on finalized
layouts for various benchmarks, including the large-scale \emph{IBM superblue} suite.  We contrast the superior resilience of our layouts with prior protection schemes, and make our
layouts publicly available,
	along with the library definition for elevating cells~\cite{webinterface}.
\end{itemize*}

\section{Background and Motivation for Our Work}
\label{sec:background_motivation}
\subsection{On Prior Studies and Some Limitations}

\textbf{Attack Schemes:}
Naive SM (i.e., splitting a layout as is) likely fails to avert skillful attackers.
That is because physical design
	tools arrange gates to be connected as close as possible, subject to available routing resources and other constraints.
Rajendran \emph{et al.\ }\cite{rajendran13_split}
introduced the concept of \emph{proximity attack} where that insight is exploited to
infer undisclosed interconnects.
More recently, Wang \emph{et al.\ }\cite{wang16_sm}
proposed a network-flow-based attack which utilizes further hints
	such as the direction of 
dangling wires and constraints on both load capacitances and delays.
Maga\~{n}a \emph{et al.\ }\cite{magana16,magana17}
	utilized routing-based proximity
in conjunction with placement-centric proximity.

\textbf{Protection Schemes:}
Various techniques have been put forward
to protect
SM-based designs
from
proximity attacks.  
Swapping of block pins
was proposed by Rajendran \emph{et al.\ }\cite{rajendran13_split}
to obtain an unbiased \emph{Hamming distance} of 50\% between the outputs of the original netlist and the outputs of the netlist restored by an attacker.
Wang \emph{et al.\ }\cite{wang16_sm} proposed
gate-level placement perturbation
within
an optimization framework, to maximize resilience and minimize wirelength overhead at the same time.
Sengupta \emph{et al.\ }\cite{sengupta17} also pursued
various placement perturbation techniques, along with a discussion on information leakage for SM.
Wang \emph{et al.\ }\cite{wang17} proposed
a routing-based protection scheme
applying wire lifting, deliberate re-routing, and VLSI test principles, all to tailor the Hamming distance towards 50\%.
Maga\~{n}a \emph{et al.\ }\cite{magana16,magana17} advocated inserting routing blockages to lift wires and, thus, to mitigate routing-centric attacks as those proposed in their study.

Besides those studies addressing proximity attacks,
Imeson \emph{et al.\ }\cite{imeson13}, Li \emph{et al.\ }\cite{li18}, and Chen \emph{et al.\ }\cite{chen2016secure}
	 focus on hardware Trojans. Patnaik \emph{et al.\ }\cite{patnaik17} pursue BEOL-centric and large-scale layout camouflaging; the authors note that their scheme is also promising in the
	context of split manufacturing.

\textbf{Limitations of Protection Schemes:}
The approach of Rajendran \emph{et al.\ }\cite{rajendran13_split}
is only applicable to hierarchical designs. More importantly, pin swapping is rather limited in practice; on average, 87\% correct connections
were reported in~\cite{rajendran13_split}.

Placement-centric schemes would ideally (re-)arrange gates randomly, thereby ``dissolving''
any hint of spatial proximity.
As this likely induces excessive PPA overheads~\cite{sengupta17,imeson13}, placement perturbation is typically applied more carefully~\cite{wang16_sm,sengupta17}.
However, as we reveal in Sec.~\ref{sec:experiments}, overly restricted perturbation can provide only a little protection, especially for splitting after higher layers.

Similar to placement-centric schemes seeking to limit PPA overheads, some routing-based schemes such as~\cite{wang17} also protect only a small subset of the design (a few nets).
These techniques are further subject to available routing resources, and re-routing may be restricted to short local detours which can be easy to resolve for an advanced attacker.
Besides, implicit re-routing by insertion of blockages~\cite{magana16,magana17} falls short of explicitly protecting selected nets and controlling the routing of wires.

\subsection{On the Trade-Offs for Cost Versus Security}
\label{sec:motivation}

It is challenging to determine the most appropriate split layer as such a decision
has direct and typically opposing impact on security and cost.
Recall that some prior art promoted to split after lower metal layers.
However, this comes at a high commercial cost for the trusted BEOL fab.
In contrast, splitting after higher layers allows for large-pitch and low-end processing setups at the BEOL fab, thus reducing cost (but possibly undermining security).
For example, considering the pitches for the 45nm node
(Table \ref{tab:pitch}),
one may prefer to split after M3 (or M6, or even M8) over splitting after M1.\footnote{Splitting after other layers
	is also possible, but considering cost and applicability we suggest that any split should occur just below the next larger
pitch. This way, the BEOL fab has to
manufacture only those larger pitches.}
Further aspects promoting higher split layers are also discussed in~\cite{xiao15}.

\begin{table}[tb]
\centering
\scriptsize
\setlength{\tabcolsep}{1.6mm}
\caption{Pitch dimensions for the metal layers in the \emph{45nm} node~\cite{rajendran13_split}.}
\smallerspacecaption
\begin{tabular}{|c|c|c|c|c|c|c|c|c|c|c|}
\hline
 \textbf{Layer} & M1 & M2 & M3 & M4 & M5 & M6 & M7 & M8 & M9 & M10 \
 \\ \hline \hline
Pitch (nm) & 130 & 140 & 140 & 280 & 280 & 280 & 800 & 800 & 1600 & 1600 \\ \hline
\end{tabular}
\label{tab:pitch}
\end{table}

\begin{figure}[tb]
\vspace{1em}
\centering
\includegraphics[width=\columnwidth]{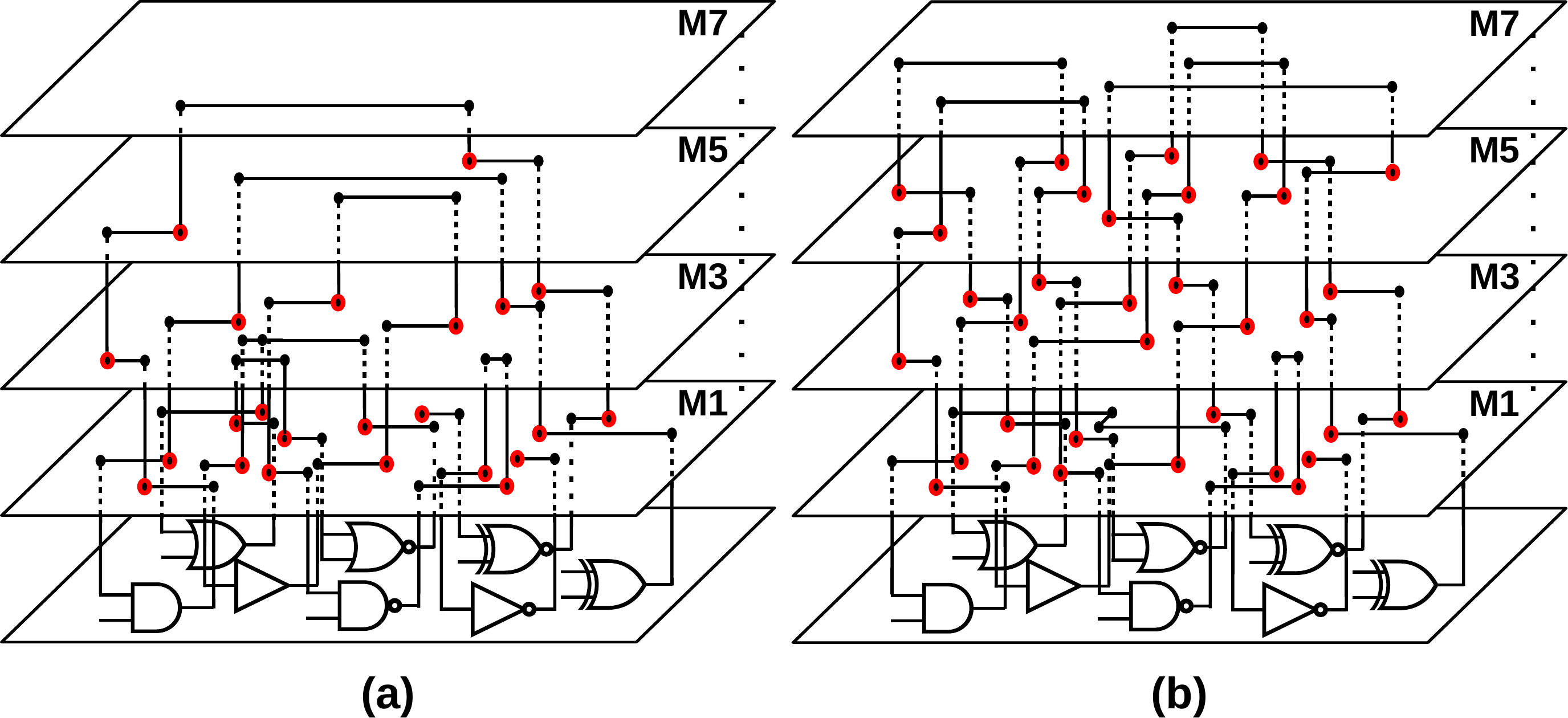}
\smallerspacecaption
\smallerspacecaption
\smallerspacecaption
\smallerspacecaption
\caption{{(a)~Conceptional illustration of a regular, unprotected layout. The red dots represent open pins, which would induce dangling wires once the layout is split at each respective layer.
Note that the majority of nets are completed in lower layers, hence fewer open pins are observed for higher layers.
(b)~Conceptional illustration of a layout protected by wire lifting.
Here the majority of nets are completed in M7 (without loss of generality). Hence, any split below M7 induces many open pins to be tackled by an attacker.}
\label{fig:SM_OPP}
}
\end{figure}

When a net is cut across FEOL and BEOL by SM, at least two \emph{dangling wires} arise in the topmost layer of the FEOL.\footnote{The reverse is not necessarily true, i.e., not all
dangling wires represent a cut net---dangling wires may also be used for obfuscation. Such wires are routed in the FEOL but remain open in the BEOL; see also
Sec.~\ref{sec:concept}.
Besides, the number of dangling wires depends both on the net's pin count and how/where exactly it is cut. See also
Fig.~\ref{fig:concept1} for an illustrative example.}
Dangling wires remain unconnected at one end;
	these
open ends indicate the locations where the vias
linking the FEOL and BEOL are to be manufactured (by the BEOL fab).
We refer to those via locations as open pins (Fig.~\ref{fig:SM_OPP}). Further, we define
\emph{open pin pairs (OPPs)} as pairs $(p_d, p_s)$
where $p_d$ is connected to a driver and $p_s$ to at least one sink. The related routing is observable in the
FEOL, but the true mapping of drivers to sinks is comprehensible only with the help of the BEOL.

\begin{figure}[tb]
\centering
\includegraphics[width=\columnwidth]{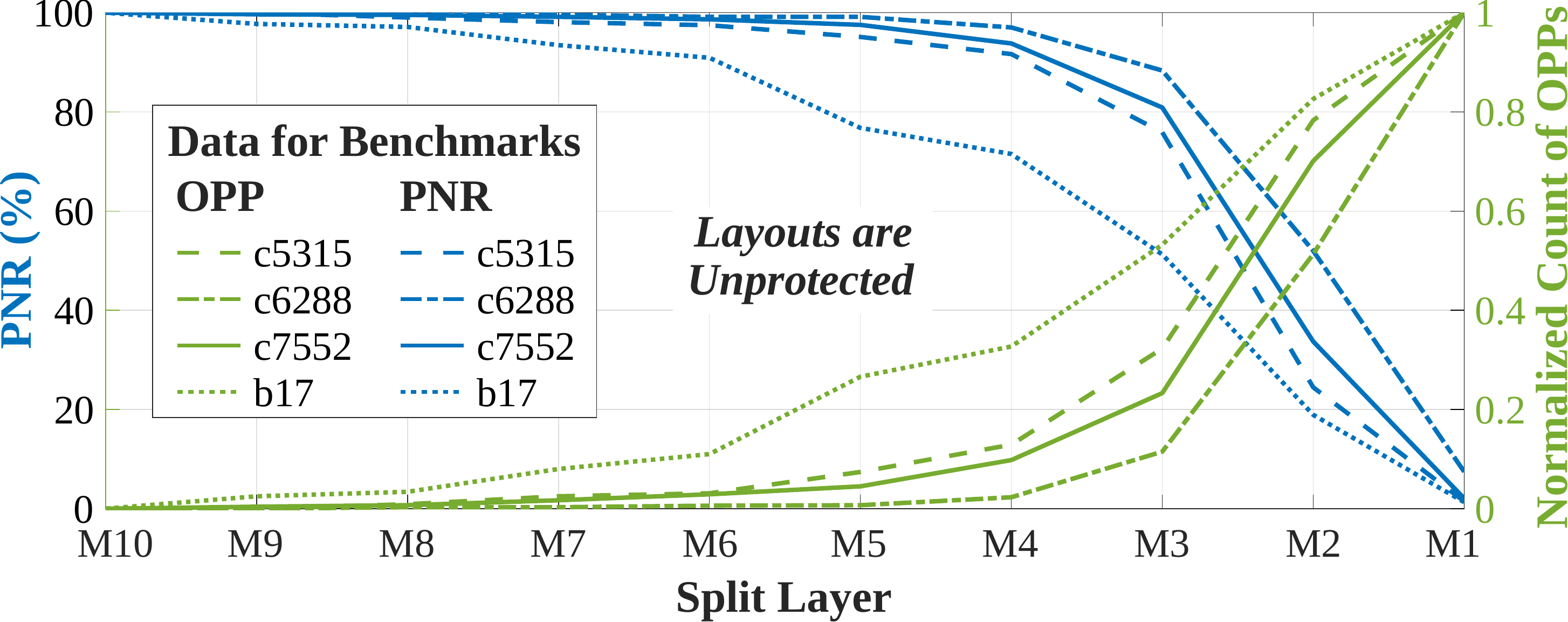}
\caption{Percentage of netlist recovery (PNR,
		see also Sec.~\ref{sec:metrics}) versus open pin pairs (OPPs), plotted as a function of
the split layer.
Note that the split layers are ordered from M10 to M1.
The unprotected layouts are naively split as is and the attack is based on~\cite{wang16_sm}.}
\label{fig:motivation}
\end{figure}

For an attacker operating at the FEOL,
observing fewer OPPs directly translates to a reduced solution space and, thus, may lower her/his efforts
for recovery of the protected design.
In Fig.~\ref{fig:motivation},
	we plot an attacker's success rate 
versus the OPP count for various split layers. There are strongly reciprocal correlations across the layers, confirming that layouts split after higher layers
	are much easier to attack.
That is because more and more nets are routed completely within the FEOL
	once we split after higher layers.
	Naturally, these FEOL-routed nets yield no OPPs and, hence, impose no efforts for the attack.

\begin{figure}[tb]
\centering
\includegraphics[width=\columnwidth]{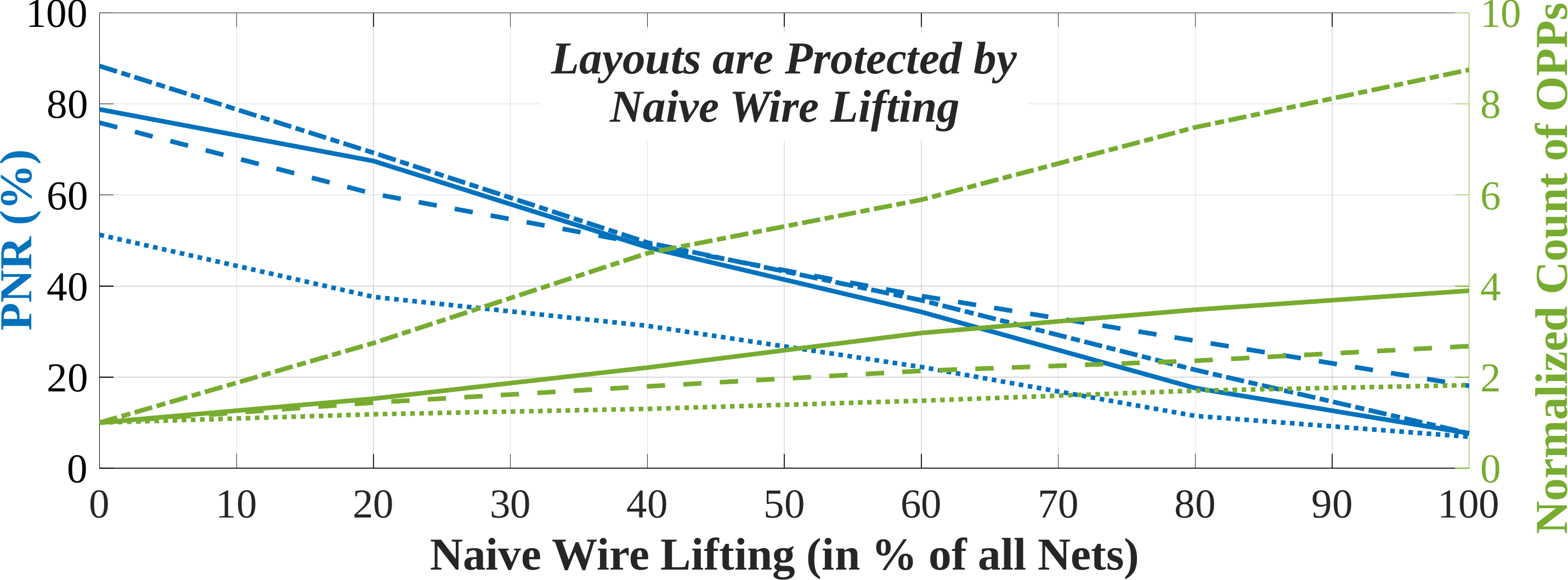}
\caption{PNR versus OPPs,
	plotted as a function of
randomly selected nets lifted to M6.
The set of benchmarks and the legend are the same as in Fig.~\ref{fig:motivation}.
The protected layouts are split after M3 and the attack is based on~\cite{wang16_sm}.
The OPP baselines (normalized OPP count of 1.0) are derived from each respectively unprotected layout, i.e., where 0\% of all nets are lifted.
}
\label{fig:PNR_naive_lifting}
\end{figure}

\begin{figure}[tb]
\centering
\includegraphics[width=\columnwidth]{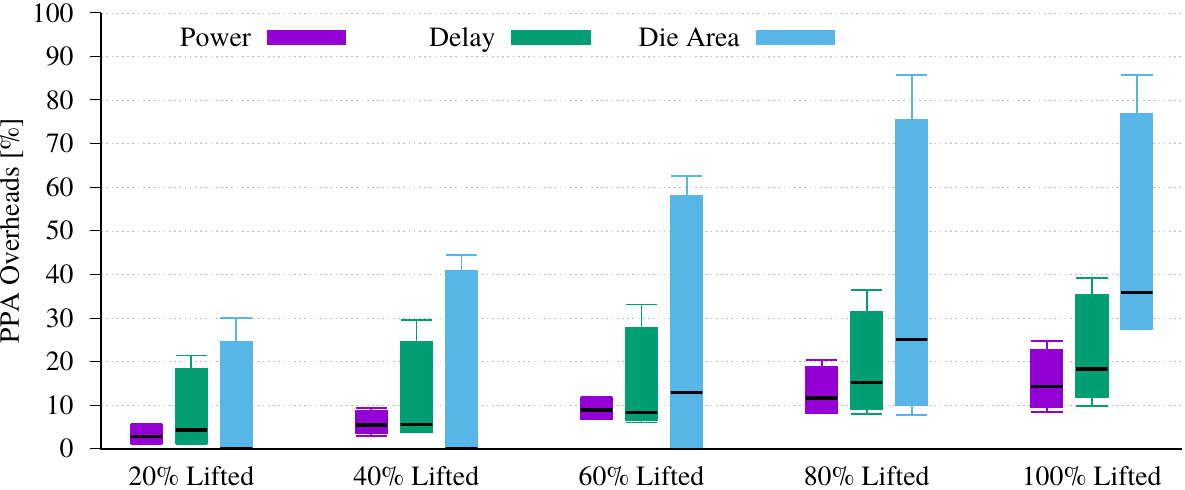}
\caption{PPA overheads for naive lifting of randomly selected nets to M6. The plot is based on the same benchmarks as in Fig.~\ref{fig:motivation}.
	The boxes span from the \nth{5} to the \nth{95} percentile, the whiskers represent the minimal and maximal values, and the black bars represent the medians, respectively.
}
\label{fig:motivation_PPA}
\end{figure}

One way to enforce many OPPs while splitting only after higher layers is \emph{wire
lifting}, i.e., the deliberate routing of nets towards and within the BEOL
(Figs.~\ref{fig:SM_OPP} and~\ref{fig:PNR_naive_lifting}).
There is a common concern of overly large PPA cost for large-scale wire lifting~\cite{glsvlsi2017,imeson13}.
We confirm this in Fig.~\ref{fig:motivation_PPA}, where we plot PPA overheads for \emph{naive lifting} of randomly selected nets.\footnote{We implement naive
	lifting by placing one ``elevating cell'' next to the driver; see Secs.~\ref{sec:concept} and~\ref{sec:methodology} for details on those cells and their use.
Such naive lifting enforces routing at least to some degree above the split layer, thereby inducing OPPs and hampering an attacker's recovery rate
(Fig.~\ref{fig:PNR_naive_lifting}). However, naive lifting cannot handle OPPs in a controlled manner.}
Note that the die area is particularly impacted.
That is because lifting more and more wires in an uncontrolled manner induces notable routing congestion which, in order to obtain DRC-clean layouts, can only be managed by enlarging the die
outlines.

Overall, there is a need for an \emph{SM scheme ensuring a large number of OPPs, while splitting after higher layers (at low commercial cost), but without inducing excessive
	PPA cost}.
We believe that such a scheme will
expedite the acceptance of SM in the industry.
In this work, we propose such a scheme through the notion of \emph{concerted wire lifting}.

\section{Strategies for Concerted Wire Lifting}
\label{sec:concept}

As motivated in Sec.~\ref{sec:background_motivation}-\ref{sec:motivation}, the number of OPPs in the FEOL should be as large as possible, but not at high cost---pertaining to commercial and
PPA cost.
We tackle this problem with the help of our custom \emph{elevating cells (ECs)}.
The key idea of routing nets through ECs is to establish pins in the metal layer of choice (above the split layer),
    thereby inducing OPPs for those nets.
(See Sec.~\ref{sec:methodology} and Fig.~\ref{fig:ECs} for implementation details.)

Next, we introduce our strategies for concerted wire lifting.
They are based on exploratory but comprehensive layout-level
experiments.
These strategies outperform naive lifting as well as recent prior art regarding security while inducing only little PPA
overhead at the same time (see Sec.~\ref{sec:experiments}).

\textbf{Strategy 1, Lifting High-Fanout Nets:} We begin by lifting high-fanout nets (HiFONs)
for two reasons: ($i$) any wrong connection made by an attacker
propagates the error to multiple locations, and ($ii$) lifting HiFONs helps introduce many OPPs.
	We define nets with two or more sinks as HiFONs.\footnote{Although large fanouts may be subject to timing-driven optimization such as
	buffering or cloning~\cite{KLMH11}, we found that on average 20--30\% of all the nets in
		the benchmarks we consider have a fanout of at least 2.
		In any case, our techniques are generic and can be readily applied for any degree of fanout.}

\begin{figure}[tb]
\centering
\includegraphics[width=\columnwidth]{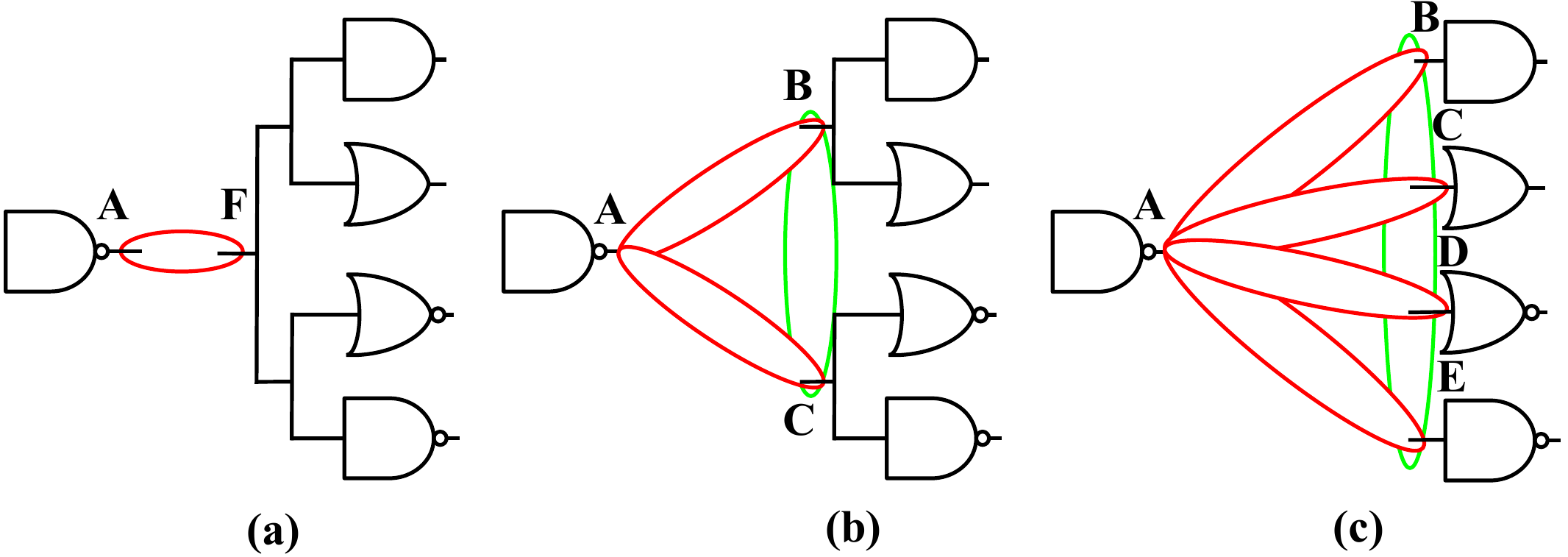}
\caption{The impact of lifting high-fanout nets on the number of open pin pairs.}
\label{fig:concept1}
\end{figure}

Consider Fig.~\ref{fig:concept1} as an example. Here,
a HiFON is originally connecting to four gates/sinks.
Depending on how and where the HiFON is lifted,
the attacker has different scenarios to cope with.
In (a), only one OPP
	arises which is trivial to attack/resolve.
In (b), two OPPs are to be tackled, (A,B) and (A,C).
Assuming that an attacker cannot tell how many sinks \emph{exactly} to consider,\footnote{While a skillful attacker may understand the driving strengths of any gate, she/he cannot easily
	resolve their original use given only the FEOL.
Any high-strength driver may have to be reconnected either to many sinks nearby or to few sinks far away.
Given that wire lifting is conducted across the whole layout, drivers and sinks of various nets will be ``spatially intermixed,'' notably increasing an attacker's efforts 
to map drivers to sinks correctly.}
either one of the two OPPs or both OPPs
at the same time are equally likely representing the original net.
Thus, the attacker has three options to consider.
In (c), even up to 14 options arise; there are four OPPs (A,B), (A,C), (A,D), and (A,E), as well as 10 possible combinations of those OPPs.
Naturally, once other nets are lifted as well, the set of OPPs
scales up even further,
   in fact in a combinatorial manner.

We lift all wires of any HiFON (Fig.~\ref{fig:concept1}(c)), to induce as many OPPs as possible.
   We do so by inserting separate ECs for the driver as well as for all the sinks.

\textbf{Strategy 2, Controlling the Distances for OPPs:}
Besides increasing the number of OPPs, it is also necessary to control the distances between their pins.
For example in Fig.~\ref{fig:concept2}(a), only a short open remains in M5 for the lifted wire/net,
motivating an attacker to reconnect that particular OPP.
Such a scenario may arise for implicit wire lifting, e.g., as proposed in~\cite{magana16,magana17}. There, only the FEOL metal layers to avoid are declared, but the actual routing
paths in the BEOL layers are not.

{\setlength\textfloatsep{8pt}
\begin{figure}[tb]
\centering
\includegraphics[width=0.80\columnwidth]{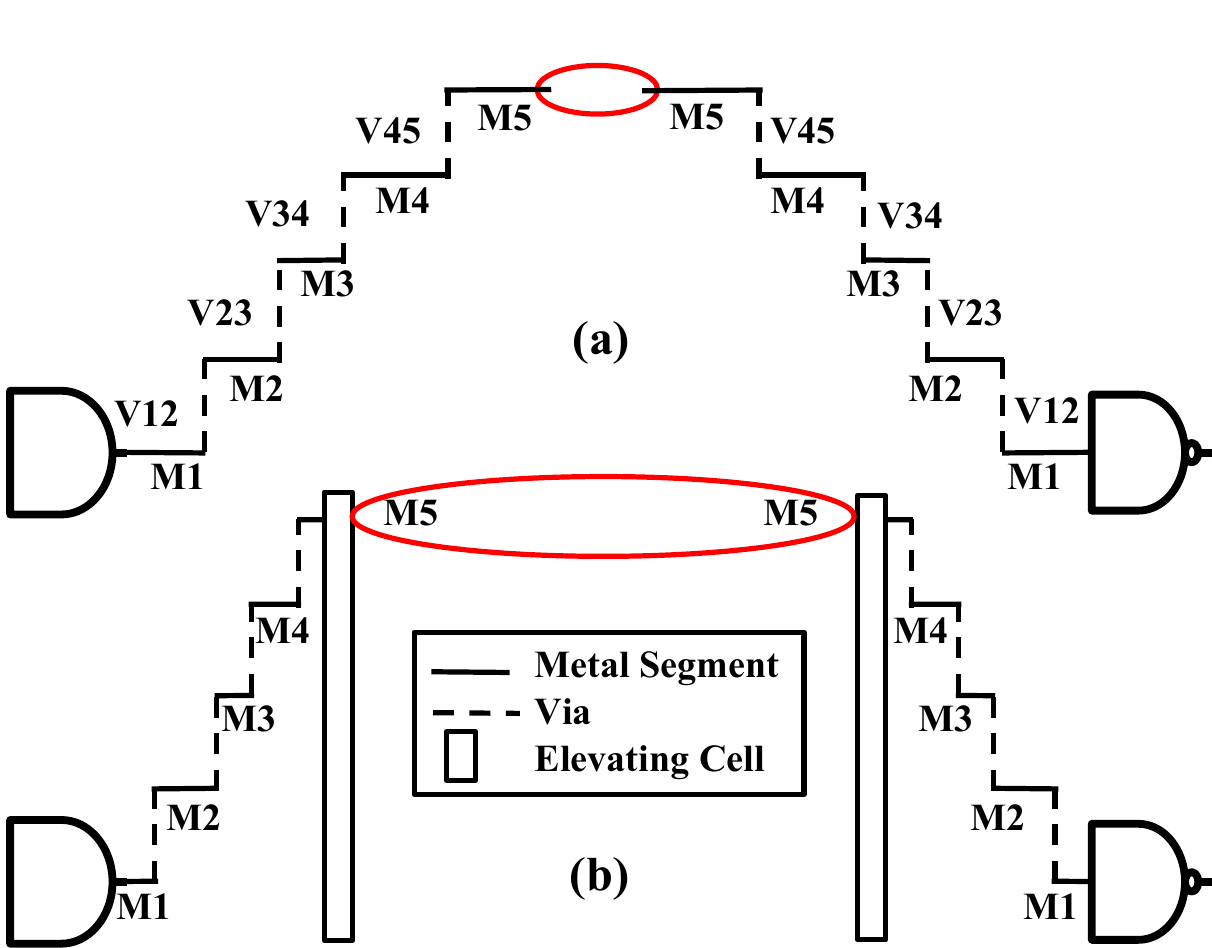}
\caption{Distances for open pin pairs (OPPs). In (a), only a short distance arises due to implicit wire lifting, whereas in (b) we precisely control the distance using two
	elevating cells (ECs).}
\label{fig:concept2}
\end{figure}
}%

In our method, we can control the distances for OPPs at will, simply by controlling the placement of the ECs (Figs.~\ref{fig:concept2}(b)). We place ECs close to the driver and the
sink(s), thereby enlarging the distances and increasing an attacker's efforts.
To mitigate any advanced attack, e.g., based on learning the distance distribution for OPPs while reverse-engineering other available chips, one may also place the ECs
	randomly within (or even beyond) the bounding boxes of the nets.

\begin{figure}[tb]
\centering
\includegraphics[width=0.92\columnwidth]{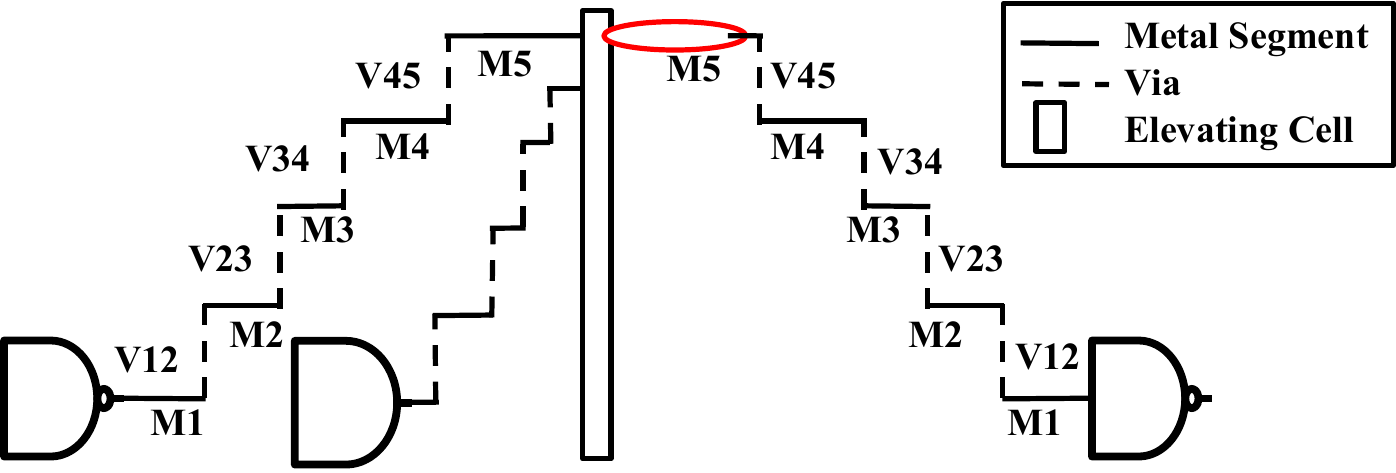}
\caption{We obfuscate short nets by connecting the
		real and dummy driver to another, dedicated type of EC.
		The dummy driver's wire connected with the EC is left dangling beyond the split layer.
		Note that the dummy driver is a real driver for another net; the related wires are not illustrated here.
			Both drivers' strengths are
			adapted such that they might seemingly connect to the sink of the obfuscated short net.
		The EC is placed between the
		real and the dummy driver, thereby increasing the OPP distances.
}
\label{fig:concept4}
\end{figure}

\textbf{Strategy 3, Obfuscation of Short Nets:}
Above we assumed that enlarging the distances of OPPs is practical and effective, which is straightforward for HiFONs (as well as for relatively long nets).
For short nets, however, enlarging those distances requires some routing detours out of the net's bounding box.
Furthermore, short nets may be easy for an attacker to identify and localize, based on the typically low driver strength.
To tackle both issues, we design another EC (Figs.~\ref{fig:concept4} and~\ref{fig:ECs}(b)).

This EC places two pins close to each other: one ``true'' pin is connected to the short net's driver, and the other ``dummy'' pin is connected to a
randomly but carefully selected
gate, representing a dummy driver.
An attacker cannot easily distinguish these two drivers:
($i$) the dummy driver is selected such that no combinatorial loops would arise were the driver connected to the short net's sink(s), and
($ii$) we adapt both drivers' strength, also accounting for the
routing detours, via ECO optimizations.
Besides obfuscation, this EC induces a \emph{dummy OPP} which naturally increases the overall number of OPPs.

Note that we insert only one EC for short nets, specifically between their real and dummy driver. We refrain from inserting another EC near the sink of short
nets, as we observed that doing so contributes little for security but hampers routability.

\section{Methodology}
\label{sec:methodology}

\begin{figure}[tb]
\centering
\includegraphics[width=.85\columnwidth]{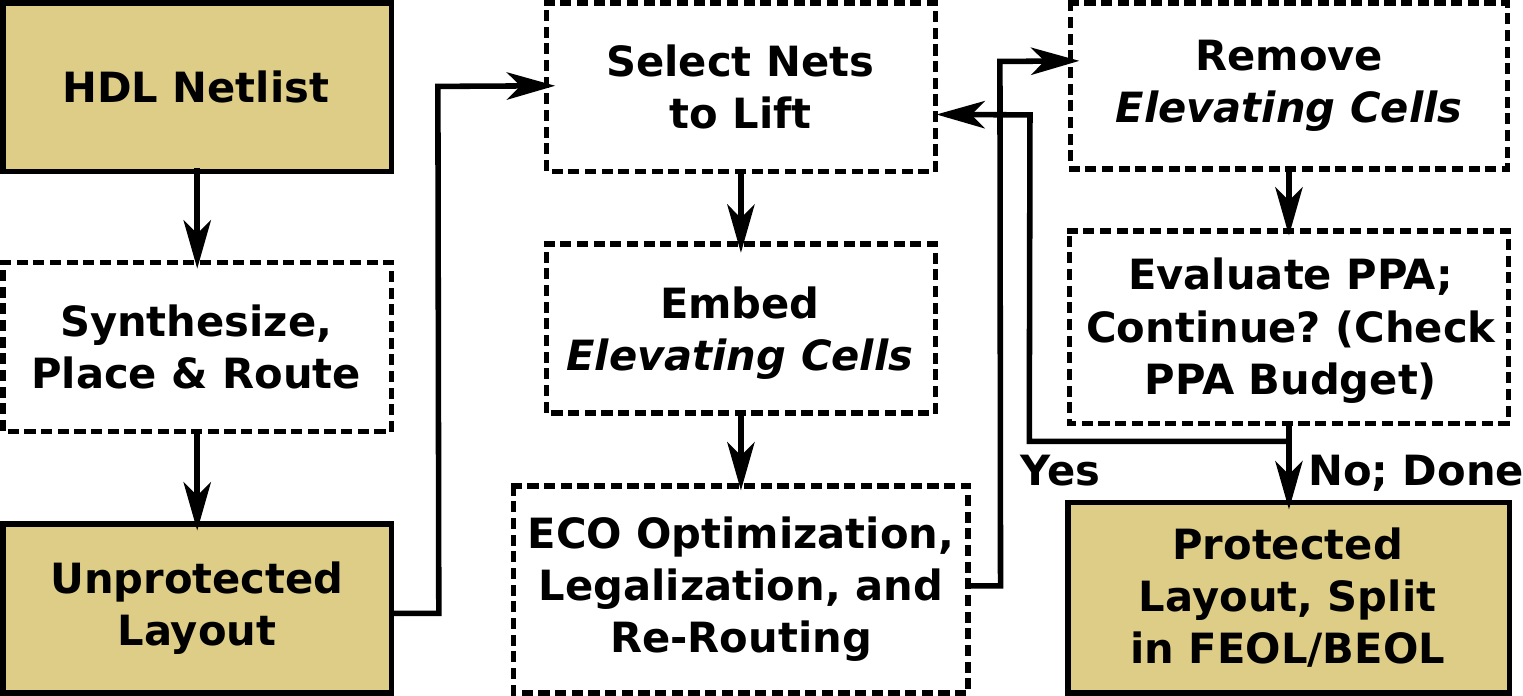}
\caption{The flow of our protection scheme.
\label{fig:flow}
}
\end{figure}

Next, we discuss our methodology
(Fig.~\ref{fig:flow}),
which is integrated with \emph{Cadence Innovus} using custom in-house scripts.
Given an HDL netlist, we first synthesize, place, and route the design. 
The resulting layout
is protected as follows.
For each net we wish to lift, \emph{elevating cells (EC)} are temporarily
inserted next to the net's driver (as well as next to all the net's sinks for HiFONs and long nets).
It is important to note that ECs do not impact the FEOL device layer; they are designed to solely elevate/lift a given net.
Next, we perform ECO optimization and legalization based on customized
scripts.
Then, we re-route the design, remove the ECs, extract the RC information, and report the PPA numbers.
In case the PPA budget allows for additional wire lifting, we continue iteratively.
Finally, a DEF file split into FEOL/BEOL is exported for security analysis against proximity attacks.

\textbf{Strategy for Selecting Nets to Lift:}
In general, we lift nets according to the strategies discussed in Sec.~\ref{sec:concept}.
More specifically, considering the iterative flow outlined above, we take 
the following steps to determine all nets to be lifted.
\begin{enumerate*}
\item Given a ratio of nets to lift,
	we initially lift HiFONs and then long nets using Strategies 1 and 2.
Here we prioritize HiFONs based on their fanout degree; large-fanout HiFONs are lifted first.
Furthermore, it is easy to see that the longer a net, the more freedom we have for controlling its OPP distance(s), and the less likely it is for an attacker to reconnect that net successfully.  
Therefore, we prioritize nets not already lifted as HiFONs by their wirelength.

\item We then lift short nets
using Strategy 3, until a given PPA budget is utilized.
We prioritize nets based on their wirelength---the shortest nets are selected first.
That is because the shorter a net, the easier it is to successfully reconnect by an attacker.
Since the additional wires required to connect with dummy drivers
consume notable routing resources, we lift short nets in small steps of 10\% and iteratively
monitor the PPA impact.
\end{enumerate*}

\begin{figure}[tb]
\centering
\includegraphics[width=.92\columnwidth]{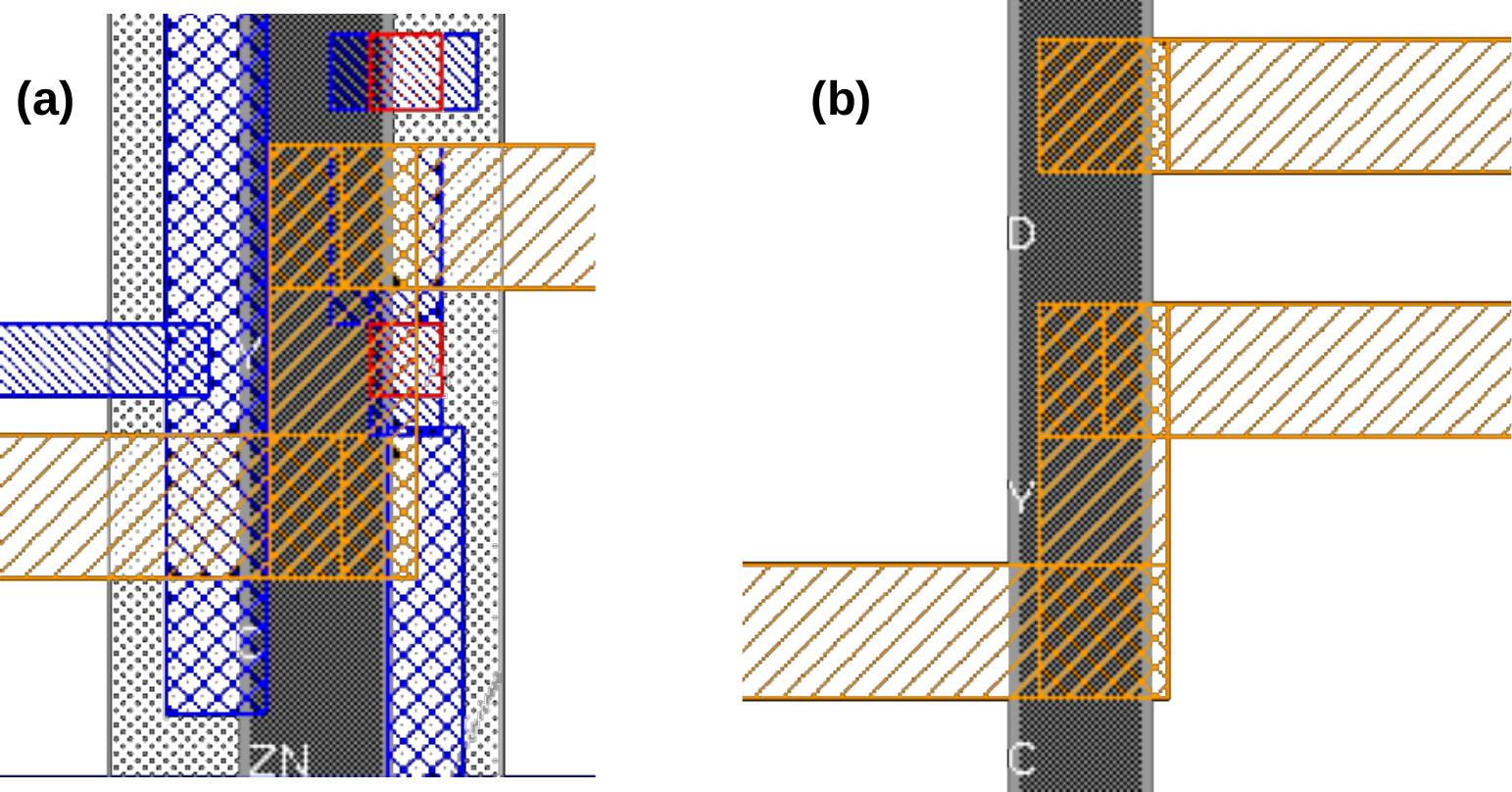}
\caption{Post-processed snapshots of our two types of ECs. Wires in M6 are in orange, wires/vias in M1 are in blued/red.
	In (a), the EC (dark grey) is seen overlapping an inverter (light grey, dotted).
		In (b), the EC is seen alone; this EC has two inputs (C, D), as required for obfuscation of short nets.
		Recall that the OPPs arise after the split layer, i.e., below M6---OPPs are not visible here.
}\label{fig:ECs}
\end{figure}

\textbf{Design of Elevating Cells:}
	  As with any custom cell,
	  our ECs are embedded in a library
	of choice.
	  We make our EC implementation for the \emph{Nangate 45nm} library publicly available in~\cite{webinterface}.
Fig.~\ref{fig:ECs} illustrates the two different types of ECs.
	  The key properties of our ECs are discussed next.

\smallerspacecaption
\begin{enumerate*}

\item All I/O pins are set up in one metal layer. Since the pins must reside above the split layer to effectuate wire lifting,
we implement different ECs as needed for various layers.

\item The pin dimensions
and offsets are chosen such that the
pins can be placed onto the respective metal layer's tracks.
This helps minimize the
routing congestion. 

\item ECs may overlap with any other standard cell (Fig~\ref{fig:ECs}(a)).
That is because the latter have their pins exclusively in lower metal layers, whereas ECs neither impact those layers nor the FEOL device layer.

\item Custom legalization scripts have been set up to prevent
the pins of different ECs to overlap with each other.

\item Timing and power characteristics
of a \emph{BUFX2} cell (buffer with driving strength 2) are leveraged
for the ECs.
A detailed library characterization is not required since ECs only translate to some interconnects in the BEOL.

\item To enable proper ECO optimization, the ECs are set up for load annotation at design time.
	That is required to capture the capacitive load of ($i$) the wire running from the EC to the
	sink and ($ii$) the sink itself. Note that this annotation is also essential for obfuscating the dummy drivers' strength as outlined in Strategy 3 (Sec.~\ref{sec:concept}).

\end{enumerate*}

\section{Metrics for Layout Protection}
\label{sec:metrics}

Here we discuss metrics to gauge the resilience of layouts when accounting for FEOL-based attacks.
First we review established metrics and then we introduce a novel metric.

The \textbf{Hamming Distance (HD)}
quantifies the average bit-level mismatch between the outputs of the original and the attacker's reconstructed design~\cite{rajendran13_split}.
Note that the HD reveals the degree of functional mismatch, but not necessarily structural mismatches.
(That is because any Boolean function can be represented by different gate-level designs.)
Hence, the HD
cannot adequately quantify the 
potential for gate-level IP theft.

The \textbf{Output Error Rate (OER)}
indicates the probability for any bit per output
being wrong while applying a possibly large set of inputs to the attacker's netlist~\cite{wang16_sm,wang17}.
As this metric tends to approach 100\% for any imperfect attack,
it does not reflect well on the degree and type of errors made by an attacker, but rather whether any error was made at all.
Like the HD, it should not be used to quantify the gate-level resilience.

The \textbf{Correct Connection Rate (CCR)} is the ratio of connections correctly inferred by an attacker over the number of protected nets.  For example, if 20 out of 100 protected nets are correctly
reconnected, the CCR is 20\%. 
Note that Wang \emph{et al.\ }\cite{wang17} defined
an \emph{Incorrect Connection Rate (ICR)}, which
is simply the inverse of this metric. 
Unlike the HD or OER, this metric can quantify the gate-level protection (or its failure).

Our metric \textbf{Percentage of Netlist Recovery (PNR)} captures the ratio of correctly inferred connections over the total number of nets.
It quantifies the structural similarity between the original netlist and the attacker's netlist. Thus, the PNR is more generic and comprehensive than the CCR, as it accounts for the
entire netlist, not only for protected nets.
Vice versa, the CCR can be considered a special case of the PNR. For unprotected layouts, both metrics shall be equal by definition.

For example, consider again that an attacker reconstructs 20 out of 100 protected nets, out of 10,000 nets in total.
Now consider further that an attacker can readily identify all nets completely routed in the FEOL.
Assuming that 2,000 nets are routed in the FEOL, the PNR would be 20.2\%.
For 6,000 nets routed in the FEOL, however, the PNR would be already 60.2\%---while the CCR remains 20\% for both cases.

In short, the PNR quantifies ($i$) the overall potential of IP theft and ($ii$) the resilience of any SM protection scheme against varyingly effective attacks, and for varying
split layers.

\section{Experimental Investigation}
\label{sec:experiments}

Recall that we propose an SM scheme enabling a large number of OPPs while splitting after higher layers, and with controllable PPA overheads.
Hence, we evaluate our scheme thoroughly regarding security as well as layout cost.

\textbf{Setup for Layout Assessment:}
Our techniques
are implemented
as custom procedures
for \emph{Cadence Innovus 16.15}.
Our procedures impose negligible runtime overheads.
We use the \emph{Nangate 45nm Open Cell Library}~\cite{nangate11};
we
utilize all ten metal layers. 
The PPA analysis has been carried out at 0.95V and 125$^{\circ}$C for the slow process corner
with a default switching activity of 0.2.
Timing results are obtained by \emph{Innovus} as well.
Our ECs lift wires to M6 unless stated otherwise.

\textbf{Setup for Security Analysis:}
We empower an attacker with the FEOL layout
	and with the technology libraries. We do not assume a working chip being available---it is yet to be
manufactured.
We utilize the network-flow-based attack provided by Wang \emph{et al.\ }\cite{wang16_sm}. Other attacks such as those in~\cite{magana16,magana17} have not been available to us at the time of
writing.\footnote{Given the focus on academic tools in~\cite{magana16,magana17}, we further assume that these attacks are not readily compatible with our industrial design flow.}
Functional equivalence was validated using \emph{Synopsys Formality}.
The OER and HD
are calculated using \emph{Synopsys VCS}
by applying 100,000 random input vectors.

\textbf{Benchmarks:} We conduct 
our comprehensive experiments
using in total 28 benchmarks, selected not only from the ``traditional'' suites (i.e., \emph{ISCAS-85}, \emph{MCNC}, and \emph{ITC-99}),
   but also from the large-scale 
\emph{IBM superblue} suite~\cite{viswanathan2011ispd}.
For the latter, we leverage scripts from~\cite{kahng14} to generate LEF/DEF files, but we also use the \emph{Nangate 45nm} library~\cite{nangate11} while doing so.

\textbf{Setup for Comparisons:} The unsplit but protected, full layouts of~\cite{wang16_sm, wang17} have been provided to us as DEF files. However, we were not made aware of
($i$) the intended split layer, ($ii$) the selection of protected nets, or ($iii$) the library files.
As for ($i$), there are indications in the layouts that they have been tailored for splitting either after
M3, M4, or M5. Hence, we calculate any comparative PNR values as average over those layers.  As a result of ($ii$), we cannot verify the other metrics but simply quote them from
the respective publications.
Because of ($iii$) we cannot contrast PPA numbers.

\textbf{Public Release:} We provide our EC implementation in~\cite{webinterface}, enabling others to protect their layouts likewise.
Moreover, we provide our final layouts as reference cases as well in~\cite{webinterface}.

\subsection{Security Analysis}
\label{sec:security_analysis}

\textbf{Increase in OPPs:}
Recall that more OPPs helps make 
proximity attacks challenging, which is corroborated by a
reduction in PNR (Figs.~\ref{fig:motivation} and~\ref{fig:PNR_naive_lifting}).
From our exploratory comparison of lifting strategies in Table~\ref{tb:open-pin-pairs} it is apparent that
our strategies 
successfully increase the 
number of OPPs over both original layouts and
layouts where naive wire lifting is employed.

As it depends on the benchmark whether the lifting of HiFONs and long nets (Strategies 1 and 2) or short nets (Strategy 3) induces more OPPs,
we suggest to apply our strategies in conjunction, as proposed in Sec.~\ref{sec:methodology}.
Next, we confirm the superior resilience of our strategies while evaluating the PNR.

\begin{table}[tb]
\centering
\scriptsize
\setlength{\tabcolsep}{1mm}
\caption{Total number of OPPs while splitting after M5.
	For a fair comparison, here we lift without loss of generality 30\% of the nets for all benchmarks and strategies.
}
\smallerspacecaption
\begin{tabular}{|c|c|c|c|c|}
\hline
\multirow{2}{*}{\textbf{Benchmark}} & \multicolumn{4}{|c|}{\textbf{OPPs}} \\
\cline{2-5}
 &
 \textbf{Original} & \textbf{Naive Lifting} & \textbf{Strategies 1 and 2} & \textbf{Strategy 3} \\ \hline \hline
c432 &  
6 & 67 & 86 & 103\\ \hline
c880 &  
8 & 116 & 170  & 204\\ \hline
c1355 & 
6 & 120 & 147 & 164\\ \hline
c1908 & 
13  & 142  & 219 & 269\\ \hline
c2670 & 
112 & 228 & 356  & 315\\ \hline
c3540 & 
53 & 386 & 576  & 458\\ \hline
c5315 & 
196 & 582 & 845 & 780\\ \hline
c6288 & 
38 & 1,235 & 1,590  & 1,630\\ \hline
c7552 & 
127 & 533 & 900 & 795\\ \hline
\end{tabular}
\label{tb:open-pin-pairs}
\end{table}

\begin{figure}[tb]
\centering
\includegraphics[width=\columnwidth]{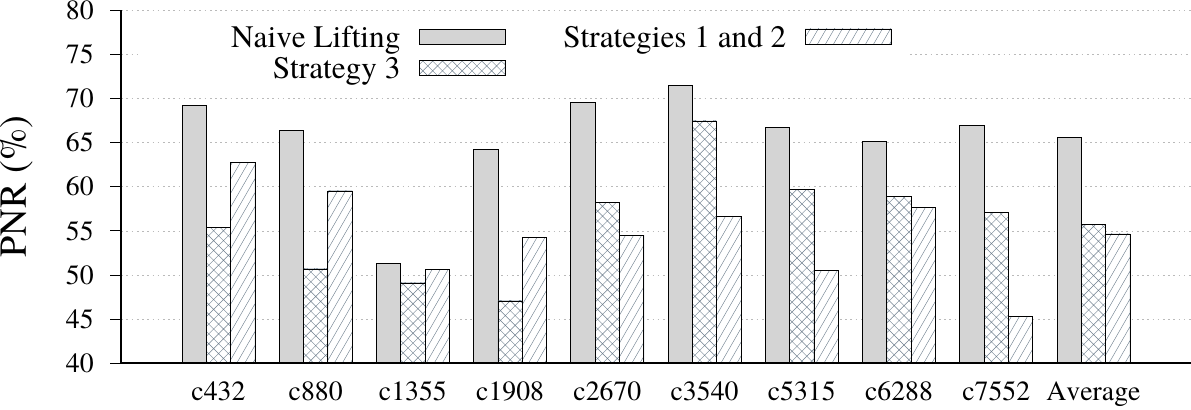}
\caption{Comparison of PNR values for naive lifting and our proposed strategies.
The attack is based on~\cite{wang16_sm}.
The data reflects the average recovery/piracy rate for splitting layouts after M3, M4, and M5.
}
\label{fig:compare1}
\smallerspacecaption
\smallerspacecaption
\end{figure}

\textbf{On the Effectiveness of Our Scheme:}
Fig.~\ref{fig:compare1} compares the PNR
for ($i$) naive lifting, ($ii$) lifting using our Strategies 1 and 2,
and ($iii$) lifting using our Strategy 3.
For a fair and comprehensive comparison, as with the exploratory comparison of induced OPPs above,
   here we lift the same percentage of nets (i.e., 30\%) for all benchmarks.
	We derive the average PNR while splitting the layouts after M3, M4, and M5.

We achieve an improvement (i.e., reduction of PNR) of 10--11\% for our strategies over naive lifting on average.
   Also here we observe that
   it depends on the benchmark which lifting technique is more effective. Thus, 
we apply our techniques in conjunction for all remaining experiments---this helps to lower PNR values even further (see below and Table~\ref{tab:metrics_compare}).

\textbf{Comparison with Prior Art:}
Initially, as a baseline comparison to the most recent work of Wang \emph{et al.\ }\cite{wang17},
   we protect the same number of nets
as they do,
   but the actual selection of nets to protect/lift is
based on our strategies.
We achieve an average improvement of $\approx$24\% for the PNR here
(Table~\ref{tb:comparison_with_JV}).

\begin{table}[tb]
\centering
\scriptsize
\setlength{\tabcolsep}{1mm}
\caption{Comparison with~\cite{wang17}
on average PNRs (over M3, M4, and M5) for the same number of nets selected for protection.
The attack is based on~\cite{wang16_sm}.}
\smallerspacecaption
\begin{tabular}{|c|c|c|c|}
\hline
\textbf{Benchmark} & \textbf{Protected Nets} & \textbf{PNR for~\cite{wang17}}
& \textbf{PNR for Proposed Scheme} \\ 
\hline \hline
c432 & 34 & 87.5 & 70.7\\ \hline
c1355 & 74 & 84.8 & 60.1\\ \hline
c1908 & 56 & 91.2 & 58.9\\ \hline
c7552 & 66 & 93.9 & 70.2\\ \hline
\end{tabular}
\label{tb:comparison_with_JV}
\end{table}

In Table~\ref{tab:metrics_compare}, we contrast the schemes of~\cite{wang16_sm, wang17} and our regular scheme, where the scope for protection/wire lifting depends on the allocated
		PPA budgets (see also Subsec.~\ref{sec:PPA_analysis}).

Naturally, original layouts without any protection are most vulnerable, 
and an attacker recovers 96\% of the netlist on average.
Constrained placement perturbation as proposed in~\cite{wang16_sm} 
provides only little improvement, reducing the average PNR to 95\%.
That is because routing eventually compensates for any gate-level perturbation, with small displacements typically being re-routed in lower metal layers (which may be readily available
		to an attacker).
The routing-centric scheme of~\cite{wang17}
can lower the PNR to 88.5\%.
In contrast, our scheme offers significantly better protection---with 31\% PNR on average, the resilience improves by
57--64\% over the prior art of~\cite{wang16_sm, wang17}.

Besides the comparison based on our PNR metric, we also contrast our scheme using established metrics (Sec.~\ref{sec:metrics}).
As for the CCR,
we note that the approach of~\cite{wang16_sm} provides again only little improvement (2.4\%) over unprotected, original layouts.
The scheme of~\cite{wang17}, however, achieves an improvement of 21.9\%, reducing the average CCR to 72.4\%.
Also here, our approach provides superior protection, by means of 0\% CCR.
Our scheme further achieves an optimal OER of 100\% (as is~\cite{wang17}, but not~\cite{wang16_sm}).
Finally, we observe an average HD of 40.3\%.
This translates to improvements of 25\% and 11\% over~\cite{wang16_sm} and~\cite{wang17}, respectively, despite the fact that we do not specifically target for
optimal HD (50\%) in our scheme.

\begin{table*}[tb]
\centering
\scriptsize
\setlength{\tabcolsep}{0.9mm}
\caption{Comparison with original layouts and prior art.
We calculate all PNR values as average over splitting layouts after M3, M4, and M5.
	CCR, OER, and HD values for original layouts, \cite{wang16_sm}, and \cite{wang17} are all quoted from \cite{wang17}; CCR is derived as (100\% - ICR).\newline
	We also report our PPA cost. All values are in percentage.
	The attack is based on~\cite{wang16_sm}.
}
\smallerspacecaption
\begin{tabular}{|*{24}{c|}}
\hline
{\textbf{Benchmark}}
& \multicolumn{4}{|c|}{\textbf{Original Layout}} 
& \multicolumn{4}{|c|}{\textbf{Placement Perturbation \cite{wang16_sm}}} 
& \multicolumn{4}{|c|}{\textbf{Routing Perturbation \cite{wang17}}} 
& \multicolumn{7}{|c|}{\textbf{Proposed Scheme}} \\
\cline{2-20}
& \textbf{\emph{PNR$^*$}} & \textbf{CCR$^*$} & \textbf{OER} & \textbf{HD} 
& \textbf{\emph{PNR}} & \textbf{CCR} & \textbf{OER} & \textbf{HD}  
& \textbf{\emph{PNR}} & \textbf{CCR} & \textbf{OER} & \textbf{HD}  
& \textbf{\emph{PNR}} & \textbf{CCR} & \textbf{OER} & \textbf{HD$^{**}$}  
& \textbf{Die-Area Cost} & \textbf{Power Cost} & \textbf{Delay Cost} \\ \hline 
\hline
c432 & 97.9 & 92.4 & 75.4 & 23.4
& 95.1 & 90.7 & 98.8 & 41.8
& 87.5 & 78.8 & 99.4 & 46.1
& 32.3 & 0 & 100 & 45.9 
& 7.7 & 13.1 & 11.6
\\ \hline

c880 & 100 & 100 & 0 & 0
& 95.3 & 96.8 & 15.8 & 1.2
& 86.8 & 47.5 & 99.9 & 18.0
& 28.3 & 0 & 100 & 39.8
& 0 & 12.1 & 19.9
\\ \hline

c1355 & 98.3 & 95.4 & 59.5 & 2.4
& 97.2 & 93.2 & 94.5 & 8.0
& 84.8 & 77.1 & 100 & 26.6
& 32.8 & 0& 100 & 46.1
&0 &12.2 & 21.3
\\ \hline

c1908 & 98.9 & 97.5 & 52.3 & 4.3
& 96.9 & 91.0 & 97.8 & 17.7 
& 91.2 & 83.8 & 100 & 38.8
& 29.5 & 0 & 100 & 48.1
&7.7 &14.6 &18.9
\\ \hline

c2670 & 89.6 & 86.3 & 99.9 & 7.0
& 94.3 & 86.3 & 100 & 7.5
& 86.3 & 58.3 & 100 & 14.0
& 34.3 & 0 & 100 & 35.1
&7.7 &10.0 &12.0
\\ \hline

c3540 & 93.8 & 88.2 & 95.4 & 18.2
& 88.5 & 82.6 & 98.8 & 27.9 
& 86.2 & 77.0 & 100 & 36.1
& 30.8 & 0 & 100 & 46.4
&7.7 &5.0 &2.8
\\ \hline

c5315 & 91.7 & 93.5 & 98.7 & 4.3
& 94.1 & 91.1 & 98.7 & 12.5 
& 87.7 & 74.7 & 100 & 18.1
& 31.6 & 0 & 100 & 35.4
&7.7 &7.9 &16.9
\\ \hline

c6288 & 97.0 & 97.8 & 36.8 & 3.0
& 95.9 & 97.6 & 74.2 & 16.5
& 92.1 & 80.9 & 100 & 42.1
& 35.6 & 0 & 100 & N/A$^{**}$
&27.3 &12.3 &15.7
\\ \hline

c7552 & 93.8 & 97.8 & 69.5 & 1.6
& 97.1 & 97.9 & 81.7 & 3.1
& 93.9 & 73.9 & 100 & 20.3
& 26.9 & 0 & 100 & 25.7
&16.7 &9.3 &15.7
\\ \hline
\hline

\textbf{Average} & 95.7 & 94.3 & 65.3 & 7.1
& 94.9 & 91.9 & 84.5 & 15.1
& 88.5 & 72.4 & 99.9 & 28.9
& 31.3 & 0 & 100 & 40.3
&9.2\% &10.7\% &15.0\%
\\ \hline

\end{tabular}
\\
$^*$ By definition, PNR and CCR values for original, unprotected layouts shall be equal. For consistency, however, we report average PNR values here as well.\\
$^{**}$ The attack of~\cite{wang16_sm} tends to provide netlists with combinatorial loops, hindering their simulation. Those netlists have been post-processed using scripts
of~\cite{kahng14}. For the benchmark \emph{c6288}, the post-processed netlist still fails simulation, due to ``UNKNOWN'' nets.
\label{tab:metrics_compare}
\end{table*}

We also seek to compare with the work of Maga\~{n}a \emph{et al.\ }\cite{magana16,magana17}.
However, having no access to their protected layouts of the \emph{IBM superblue} benchmarks, we can only compare on a qualitative level.
In Table~\ref{tab:comparison_for_superblue}, we contrast
their and our counts of {\emph{additional vias} above their assumed split layer, i.e., M4, and up to M6, where we lift wires to in our scheme.
Note that only the total via counts across all layers before lifting and the layer-wise differences in via counts after lifting are given in~\cite{magana16}, but not the original via counts per layer.
Considering the respective total via counts before lifting as independent baselines,
our scheme increases the vias for V45 and V56 by 2.25--3.71\% (with respect to total vias), whereas Maga\~{n}a \emph{et al.\ }increase those vias counts only by 0.67--2.03\%.

In their recent study~\cite{magana17}, Maga\~{n}a \emph{et al.\ }also report on the relative vias increases per layer; we contrast their increases with ours in
	Table~\ref{tab:comparison_for_superblue_TVLSI}.  We observe on average 74\% and 101\% more vias for V45 and V56, respectively, while the respective increases reported
		in~\cite{magana17} are roughly only 16\% and 49\%.  Note that we achieve the underlying wire lifting while keeping the die area fixed as
		in~\cite{magana17}, i.e., we induce zero area cost (and only marginal power and delay overheads, see also Subsec.~\ref{sec:PPA_analysis}).

Any increase of vias above the split layer is a direct indication of more nets being routed in the BEOL, hence inducing more OPPs and a higher complexity for proximity
	attacks. Therefore, we believe that our scheme generally renders the \emph{IBM superblue} benchmarks more
resilient.

\begin{table*}[tb]
\centering
\scriptsize
\setlength{\tabcolsep}{0.1em}
\caption{Comparison with~\cite{magana16}.
		For fair comparison, we also allow no die-area overhead.
	We report PPA numbers on DRC-clean layouts.
}
\smallerspacecaption
\begin{tabular}{|*{12}{c|}}
\hline
\multicolumn{3}{|c|}{\textbf{Benchmarks}}
& \multicolumn{3}{|c|}{\textbf{Implicit Wire Lifting~\cite{magana16}}}
& \multicolumn{6}{|c|}{\textbf{Proposed Scheme}} \\
\hline
{\textbf{Name}}
& \textbf{ Nets$^*$} & \textbf{Placement}
& \textbf{Total Vias$^*$} & \textbf{$\Delta_+$V45} & \textbf{$\Delta_+$V56}
& \textbf{Total Vias$^*$} & \textbf{$\Delta_+$V45} & \textbf{$\Delta_+$V56}
& \textbf{Power (mW)}
& \textbf{Delay (ns)}
& \textbf{Die Area $(\mu m^2)$} \\

& & \textbf{Util. (\%)}
& \textbf{Before Lifting} & \textbf{After Lifting} & \textbf{After Lifting}
& \textbf{Before Lifting} & \textbf{After Lifting} & \textbf{After Lifting}
& \textbf{After / Before}
& \textbf{After / Before}
& \textbf{After = Before} \\

\hline
\hline
\emph{superblue1}
& 879,168 & 69
& 4,597,616
& 40,051 (0.87\%)
& 70,355 (1.53\%)
& 6,679,733
& 247,739 (3.71\%)
& 233,749 (3.50\%)
& 82.7 / 81.9 & 29.8 / 29.4 & 1,520,868
\\ \hline

\emph{superblue5} & 764,445 & 77
& 4,650,756
& 34,828 (0.75\%)
& 62,704 (1.35\%)
& 5,523,805
& 139,900 (2.53\%)
& 133,052 (2.41\%)
& 79.2 / 78.6 & 24.7 / 24.6 & 1,298,221
\\ \hline

\emph{superblue10} & 1,158,282 & 75
& 6,304,110
& 42,210 (0.67\%)
& 50,999 (0.81\%)
& 8,875,439
& 228,454 (2.57\%)
& 220,176 (2.48\%)
& 116.3 / 115.5 & 29.8 / 29.5 & 2,176,080
\\ \hline

\emph{superblue12} & 1,523,108 & 56
& 8,913,075
& 151,018 (1.69\%)
& 175,614 (1.97\%)
& 11,813,683
& 265,992 (2.25\%)
& 274,908 (2.33\%)
& 127.2 / 126.3 & 28.9 / 28.8 & 2,276,426
\\ \hline

\emph{superblue18} & 672,084 & 67
& 3,582,687
& 45,417 (1.27\%)
& 72,897 (2.03\%)
& 4,852,381
& 164,971 (3.40\%)
& 163,412 (3.37\%)
& 81.3 / 80.4 & 19.7 / 19.5 & 1,158,182
\\ \hline

\end{tabular}
\\
$^*$ Values are different as we use \emph{Cadence Innovus} whereas Maga\~{n}a \emph{et al.\ }\cite{magana16} employ academic tools.
	Moreover, the metal layer corresponding to M10 in the \emph{Nangate 45nm} library~\cite{nangate11} is missing for~\cite{magana16}.
As the contribution for overall routing tracks from M10 is only 1.41\%,
	the comparison can be considered fair nevertheless.

\label{tab:comparison_for_superblue}
\smallerspacecaption
\smallerspacecaption
\smallerspacecaption

\end{table*}

\begin{table}[tb]
\centering
\scriptsize
\setlength{\tabcolsep}{0.17em}
\caption{Comparison with~\cite{magana17}. Note that our layouts are the same as for Table~\ref{tab:comparison_for_superblue}.$^*$
}
\smallerspacecaption
\begin{tabular}{|c|c|c|c|c|c|}
\hline
\multirow{3}{*}{\textbf{Benchmark}} & \multicolumn{2}{|c|}{\textbf{Implicit Wire Lifting~\cite{magana17}}} & \multicolumn{3}{|c|}{\textbf{Proposed Scheme}} \\
\cline{2-6}
 &
 \textbf{$\Delta_+$V45 (\%)} &
 \textbf{$\Delta_+$V56 (\%)} &
 \textbf{Wire Lifting} &
 \textbf{$\Delta_+$V45 (\%)} &
 \textbf{$\Delta_+$V56 (\%)} \\
	 & & & \textbf{(in \% of Nets)} & & \\
 \hline \hline
\emph{superblue1} &  
19.18 & 80.10 & 7 & 101.20 & 133.55 \\ \hline
\emph{superblue5} &  
9.43 & 29.84 & 5 & 57.53  & 76.33 \\ \hline
\emph{superblue10} &  
26.05 & 54.18 & 5 & 63.97 & 81.93 \\ \hline
\emph{superblue12} &  
9.72  & 32.40 & 5 & 55.73 & 80.27 \\ \hline
\emph{superblue18} &  
14.32 & 47.41 & 8 & 91.84  & 135.27 \\ \hline
\hline
\textbf{Average} & 
15.74 & 48.79 & 6 & 74.05 & 101.47 \\ \hline
\end{tabular}
\\
$^*$ This implies that the percentage of lifted wires is the same in Table~\ref{tab:comparison_for_superblue}. We simply report it here due to lack of space in
	Table~\ref{tab:comparison_for_superblue}.

\label{tab:comparison_for_superblue_TVLSI}
\smallerspacecaption
\smallerspacecaption
\smallerspacecaption

\end{table}

\subsection{PPA Analysis}
\label{sec:PPA_analysis}

Recall that we cannot directly compare
to the works of Wang \emph{et al.\ }\cite{wang16_sm, wang17} (and Maga\~{n}a \emph{et al.\ }\cite{magana16,magana17}).
That is because we have no access to the library (and DEF) files, and PPA cost are not reported in the respective publications.
As for our qualitative comparison with 
Maga\~{n}a \emph{et al.\ }\cite{magana16,magana17}, we also report our PPA numbers on the large-scale \emph{IBM superblue} benchmarks (Table~\ref{tab:comparison_for_superblue}).
Notably, we observe only 0.85\%, 0.83\%, and 0\% overheads for power, delay, and die area, respectively.

We next discuss in detail the PPA cost as incurred for the comparative experiments (Subsec.~\ref{sec:security_analysis}, Table~\ref{tab:metrics_compare}).
Empirically, we allow for different PPA budgets since large benchmarks such as {\em
	c6288} require more die area to maintain DRC-fixable layouts (and reasonably low PNR values).
	The average budgets for the experiments in Table~\ref{tab:metrics_compare} are 10\% for power and die area, and 15\% for delay, respectively.
		Using our flow and given these budgets, we can lift on average 50--60\% of all nets.
			This ratio of lifted nets over PPA budgets is reasonable---that is especially true in contrast to naive lifting (Fig.~\ref{fig:motivation_PPA}).

\textbf{On Area:}
Recall that our elevating cells
do not impact the FEOL area.
Besides, we initially set the utilization targets such that
less than 1\% routing congestion can be obtained.
Whenever required to enable lifting, we stepwise increase die outlines, which is then reported as die-area cost.

\textbf{On Power and Performance:} As we move
selected
nets to higher metal layers, an increase of wirelength is expected.
As a result, we also observe average overheads of 10.7\% and 15.0\% for power and delays, respectively.
One can attribute those reasonable overheads to the relatively low resistance of higher layers.
Once more and more
nets are lifted, however, that positive effect is offset by
a steady increase of routing congestion. Typically,
congestion is
managed by re-routing,
which lengthens nets further, aggravating the overheads further to some degree.
Besides, we conservatively estimate the impact of dummy OPPs.
	That is because we consider the
	annotated load of ECs, capturing the wires and sink,
	whereas only the capacitance of the dangling wire has to be driven in reality.
	
\begin{table}[tb]
\centering
\scriptsize
\setlength{\tabcolsep}{1mm}
\caption{PPA cost and PNR when using two additional metal layers.
	The attack is based on~\cite{wang16_sm}.}
\smallerspacecaption
\begin{tabular}{|c|c|c|c|c|}
\hline
 \textbf{Benchmark} &
\textbf{\em{PNR}} & \textbf{Die-Area Cost} 
& \textbf{Power Cost} & \textbf{Delay Cost} \\ \hline \hline
c5315 &  
28.1 & 0 & 2.9 & 3.3\\ \hline
c6288 &  
34.5 & 0 & 7.2  & 5.6\\ \hline
c7552 & 
24.6 & 0 & 3.5 & 4.3\\ \hline
\hline
\textbf{Average} & 
29.1 & 0 & 4.5 & 4.4\\ \hline
\end{tabular}
\label{tb:extra-metal-layers}
\smallerspacecaption
\smallerspacecaption
\smallerspacecaption
\end{table}	

\textbf{On the Use of Additional Metal Layers:}
Finally, we observe that the PPA cost (and PNR) can be further improved once
additional metal layers are employed (Table~\ref{tb:extra-metal-layers}).
Here we duplicate M6 two times, resulting in 12 layers in total.
Also, here we focus on relatively large and challenging benchmarks.

Additional metal layers can even provide a commercial benefit for SM and wire lifting, as long as higher layers are used.
That is because the relatively low mask and manufacturing cost of large-pitch, higher layers may be more than compensated for by 
the achievement of zero cost for die area---this reduces the overall footprint of SM on commercial cost significantly.

\section{Conclusion}
\label{sec:conclusion}

We propose a BEOL-centric scheme towards concerted wire lifting, advancing
the prospects of split manufacturing (SM).
Besides, our novel PNR metric helps to properly quantify the resilience against gate-level theft of intellectual property (IP).

The objectives we addressed here are ($i$)~to enable splits after higher metal layers, thereby reducing the commercial footprint of SM, ($ii$)~superior resilience, and
($iii$)~reasonable and controllable PPA cost.
We believe that schemes like ours are essential to expedite the acceptance
of SM in the industry.

We demonstrated exhaustively that our concerted lifting scheme is more effective and efficient than naive lifting, both in terms of protection and PPA cost. In our
	comparative analysis, we also found that our scheme excels prior art. For example, we achieve 0\% CCR for commonly considered benchmarks (selected from \emph{ISCAS-85},
			\emph{MCNC}, and \emph{ITC-99} suites), whereas some prior art experiences CCR well above
	70\%.
Besides 0\% CCR, we enable PNR as low as 31\% on average.
This directly translates to much better IP protection than prior art, which tends to experience 89\% PNR or even
more. Note that we may further reduce the PNR by lifting more wires, at least once higher PPA budgets
are considered as acceptable.

For future work, besides employing other upcoming proximity attacks, we will evaluate the resilience of our scheme within a formal security model.
	Also, we will further study the prospects of additional higher metal layers.

\section*{Acknowledgements}
The authors are grateful to Yujie Wang, Tri Cao and Jeyavijayan (JV) Rajendran (Texas A\&M University)
for providing their network-flow attack and their protected layouts of~\cite{wang16_sm, wang17}.
Further, the authors thank Jonathon Maga\~{n}a and Azadeh Davoodi (University of Wisconsin--Madison) for discussion.

\footnotesize
\newcommand{\BIBdecl}{\setlength{\itemsep}{-0.14em}}
\bibliographystyle{IEEEtran}
\bibliography{main}

\end{document}